\documentclass{sistedes}
\usepackage[T1]{fontenc}
\usepackage{graphicx}
\usepackage{multirow} % To combine table rows & columns
\usepackage{pdflscape} % To rotate pages
\usepackage{hyperref} % To enable hyperlinks. 
\usepackage{xcolor} % To use colors
\usepackage{subcaption} % To group images
\usepackage{amssymb} % To extend the symbol library
\usepackage{pifont}
\usepackage{tcolorbox}
\usepackage{xspace}
\usepackage[colorinlistoftodos,textsize=small]{todonotes} % To include notes in the margin
\usepackage{tabularx} % To change table columns widths

\newcolumntype{C}{>{\centering\arraybackslash}X}
\newcolumntype{L}{>{\raggedright\arraybackslash}X}

\begin{document}
\title{Pricing-driven Development and Operation of SaaS : Challenges and Opportunities
    %\thanks{Supported by organization x.}
}
%
%\titlerunning{Abbreviated paper title}
% If the paper title is too long for the running head, you can set
% an abbreviated paper title here
%
\author{Alejandro García-Fernández\inst{1}\orcidID{0009-0000-0353-8891} \and
José Antonio Parejo\inst{1}\orcidID{0000-0002-4708-4606} \and
Antonio Ruiz-Cortés\inst{1}\orcidID{0000-0001-9827-1834}}

\titlerunning{Pricing-driven SaaS DevOp: Challenges and Opportunities}

\authorrunning{García-Fernández et al.}
% First names are abbreviated in the running head.
% If there are more than two authors, 'et al.' is used.
%
\institute{SCORE Lab, I3US Institute, Universidad de Sevilla, Espa\~na \\
\email{\{agarcia29,japarejo,aruiz\}@us.es}}
\maketitle              % typeset the header of the contribution
\begin{abstract}
As the Software as a Service (SaaS) paradigm continues to reshape the software industry, a nuanced understanding of its operational dynamics becomes increasingly crucial. This paper delves into the intricate relationship between pricing strategies and software development within the SaaS model. Using PetClinic as a case study, we explore the implications of a Pricing-driven Development and Operation approach of SaaS systems, highlighting the delicate balance between business-driven decision-making and technical implementation challenges, shedding light on how pricing plans can shape software features and deployment. Our discussion aims to provide strategic insights for the community to navigate the complexities of this integrated approach, fostering a better alignment between business models and technological capabilities for effective cloud-based services.

\end{abstract}
\section{Introduction}
% Definir qué es un SaaS, y cómo su uso y la capacidad para desplegarlos en la nube ha generado oportunidades sin precedentes en el mundo del desarrollo software.
%\comentario{}{SaaS ha incrementado la necesidad de hacer software flexible.}

Software as a Service (SaaS) represents a significant shift in the software industry, delivering software over the Internet to distributed organisations and users around the globe \cite{Yao2017Explaining}. The advent of the SaaS model along with cloud-based deployment, has 
%dramatically 
influenced how software is developed, distributed, and implemented, reshaping the IT industry and its associated business ecosystems \cite{Nieuwenhuis2017}. This shift has increased the flexibility, scalability, and economical viability of software development, creating unprecedented opportunities for developers and organisations.
% Establecimiento de planes de precios con diversos planes alternativos, y subscripciones por parte de los usaurios para el uso del SaaS se ha convertido en el mecanismo de motenización más popular en este tipo de sistemas. 

% Este modelo tiene ventajas tanto para proveedores como para clientes, pero principalmente, permite adaptarse a las necesidades y patrón de uso de los usuarios.

%\comentario{}{razon por la que los prcing plans son el mecanismo más polular.}
The establishment of pricing plans (henceforth pricings)  to segment features, and the adoption of subscription models by users, have become the most popular software licensing mechanism in SaaS systems, due to their ability to provide a predictable revenue stream for providers, while offering flexibility and scalability to users \cite{Jiang2009}. 

%\comentario{}{los pricings suoponen un cambio del modelo de licneciamiento de software.}
This approach, typically characterised by usage-based charging plans, allows SaaS systems meet a diverse range of client needs. Its adoption has been pivotal in the growth and sustainability of SaaS development, representing a shift from traditional software licensing to a more dynamic and user-centric approach.
% Sin embargo, este cóctel de aproximaciones y tecnologías que denominaremos "desarrollo y operación de SaaS con despliegue en la nube dirigidos" por planes de precios, presenta también algunos desafíos muy importantes tanto para desarrolladores como para operadones que gestionas el despliegue y la operación en la nube de dichos sistemas.

%\comentario{}{se propone acuñar el PD devop of SaaS como un proceso. -- no se da definición directa.}
However, anchoring the SaaS business model on dynamic pricing strategies means that market forces frequently shape the pricing structure, such as the addition of features, alteration of usage limits, and revisions to the structure and tiers of pricing plans. To sustain competitiveness and ensure pricing plans are effectively met, those changes must be timely implemented by the developers and properly supported by the cloud infrastructure and architecture of the system. We coin this process —wherein business decisions influence pricing modifications, thereby initiating development efforts and changes on the architecture or deployment infrastructure— as the \emph{Pricing-driven development and operation of SaaS}.

% En este artículo enumeramos los que consideramos son los mayores desafíos de esta aproximación al desarrollo y operación del software, usando una aplicación típica como running example que permita ilustrar y comprender mejor la casuística, gravedad e implicaciones de dichos problemas y desafíos.
In this paper we delineate some of the primary challenges associated with the  Pricing-driven development and operation of SaaS on the Cloud. Using a typical application as running example, we aim to elucidate 
%the nuances, severity, and 
the implications of the issues that these challenges can raise. Additionally, we envision some opportunities that are enabled by future solutions to those challenges.

The remainder of this paper is structured as follows: section \ref{sec:background} define the key terms to establish a common understanding for the rest of the content and presents PetClinic as our running example, whose pricing is used to illustrate the complexity of managing pricing-driven feature toggles in section \ref{sec:limitations}; after that, the challenges and opportunities of \emph{Pricing-driven development and operation of SaaS 
%on the Cloud
} are presented in sections \ref{sec:challenges} and \ref{sec:opportunities} respectively; finally, we summarise the content of the paper in section \ref{sec:conclusion} and open the topic for discussion.
\vspace{-2pt}
\section{Background and related work}
\label{sec:background}

\paragraph{Pricing plans.} Although there is not yet a commonly accepted model that represents \emph{the pricing of a SaaS}, most providers share similar structures with multiple common elements. In this context, a pricing is a container that stores features, which are usually grouped into \emph{plans} and \emph{add-ons}. The main difference between them is that while users can only subscribe to one single plan, add-ons do not share this restriction, and users can subscribe to as many as they want. Within this model, a customer interacts with the SaaS by establishing \emph{a subscription}, through which he commits to pay a periodic fee (aka usage tariff \cite{LI2022}) to gain the ability to access and leverage the functionality and information provided by the SaaS in the terms and limits set out by the chosen plan or/and set of add-ons.

\begin{figure}
    \centering
    \includegraphics[width=\textwidth]{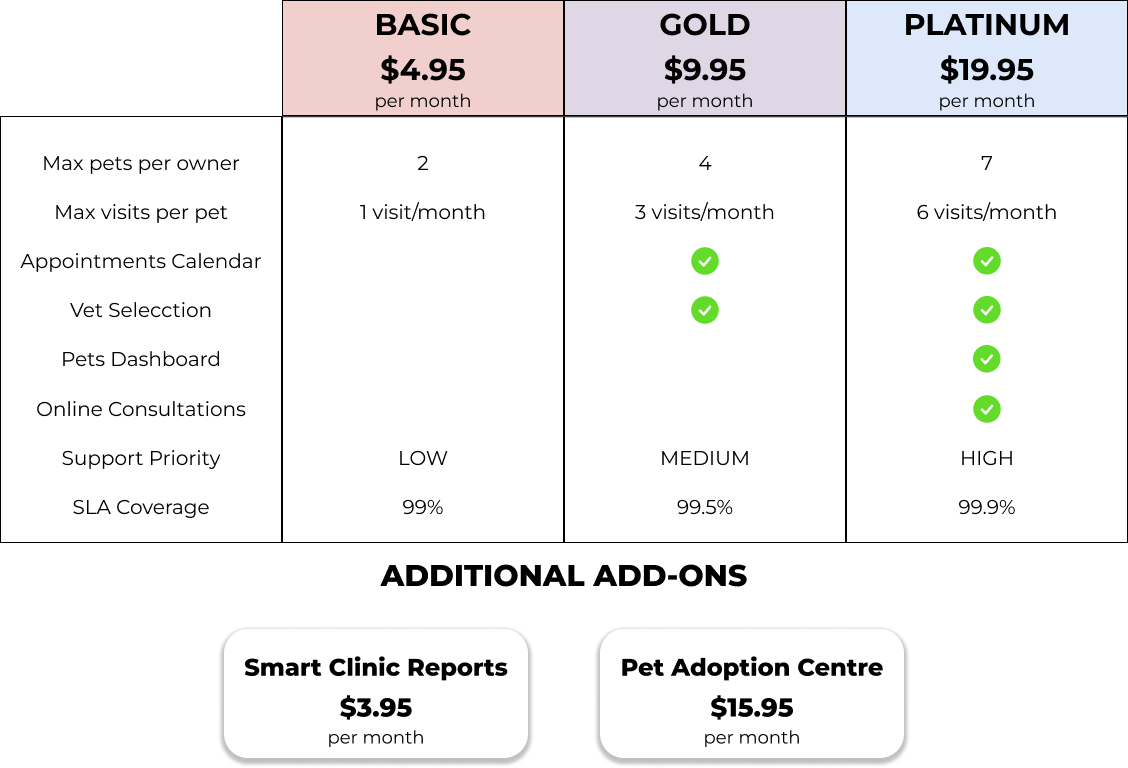}
    \caption{A potential pricing for Petclinic with 3 plans, 8 features and two add-ons. }
    \label{fig:petclinicPricing}
\end{figure}

Figure \ref{fig:petclinicPricing} illustrates a  pricing  for PetClinic, a service aimed at veterinary clinics and pet owners, offering three plans across eight features, along with two additional add-ons.
%\aruiz{La fig X muestra un posible pricing para el SaaS "Petclinic", un servicio orientado a clinicas veterinarias y usuarios de mascotas \cite{}. El plan ofrece 3 planes sobre 8 característica y también dos additional add-ons.} 
In this context, a \emph{feature} is a distinctive characteristic whose presence/absence may guide a user’s decision towards a particular subscription, e.g., "selection vet".
It is worth noting that this definition of feature involves a broader scope than the one given for SPLs \cite{10.1007/11431855_34} or feature toggling \cite{FOWLER2023,jezequel2022feature}. %\agarcia{Este comentario tras la coma, queda ya un poco desligado del hilo argumental, no?} \aruiz{si. un poco desligado. he añadido algo a los pricinhs.imho es mejor }.

\paragraph{Feature toggling.} To enable the access to different features by different users software developers use \emph{feature toggles}\footnote{For the sake of brevity, in this paper, the term feature toggle and toggle will be used interchangeably.}  a.k.a. feature flags, a powerful technique that allow to modify a system's behaviour without changing the code \cite{FOWLER2023}. The concept of feature in this context differs from pricing, conceiving a feature as a code fragment that is executed or not depending on a boolean evaluation. In the literature, this evaluation is defined as ``the process through which the state of each feature toggle is valued based on the current system's configuration'' \cite{Rosu2023} but, in SaaS platforms,  user's subscription and plan constraints on features usage can also take part in this process. 

However, despite their potential, \cite{TERNAVA2022} demonstrated, after the analysis of 5 different systems, that, on average, the 27\% of them tend to fall into disuse as the system evolves, generating ``one of the worst kinds of technical debt'', and transforming the system's testing into a combinatorial problem, given the state possibilities of toggles \cite{Rahman2016}. That's why some heuristics have been proposed to structure feature toggles efficiently, and reduce the associated debt \cite{MAHDAVI2022}.

\begin{figure}[htb]
    \centering
    \includegraphics[width=\textwidth]{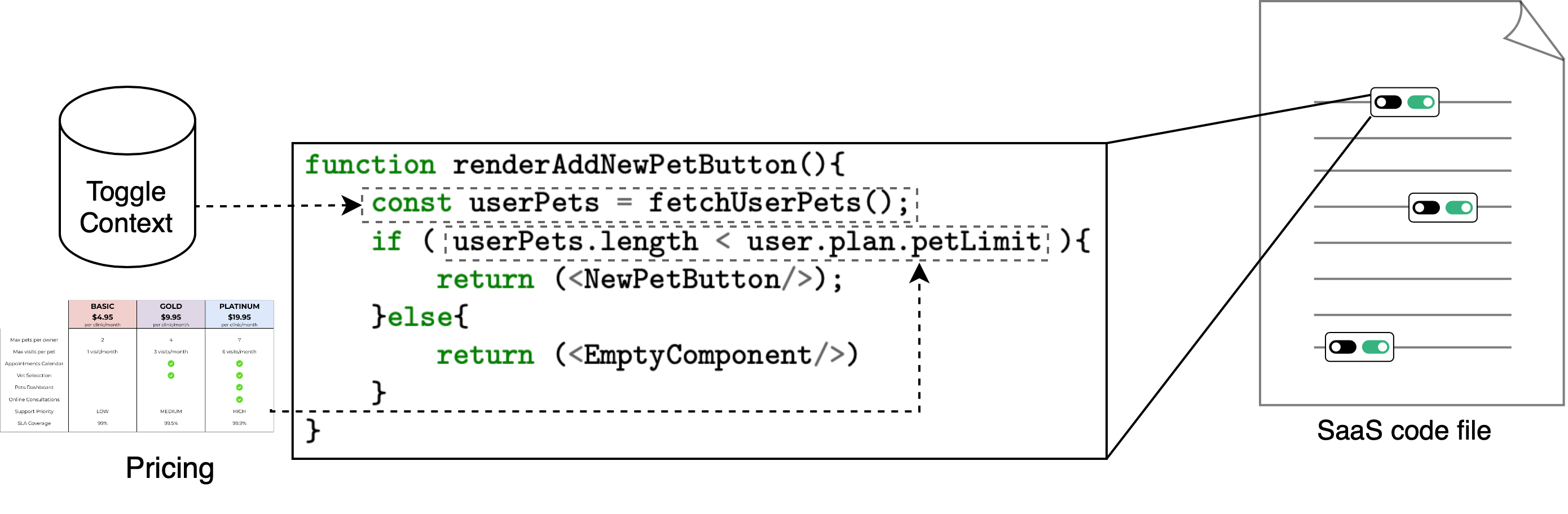}
    \caption{Feature point implementation inside a SaaS codebase}
    \label{fig:togglingExample}
\end{figure}

In Figure \ref{fig:togglingExample}, the key concepts of feature toggling are illustrated. As can be seen, a file of the code base contains \emph{toggle points}, which are locations within the code where logical condition evaluations take place, resulting in the alteration of the system's behaviour. Inside a toggle point, an evaluation to determine whether a feature is enabled or not takes place, for instance, evaluating if the number of pets already registered by the owner (userPets.length) has (not) reached the limit set in the plan (user.plan.petLimit). % \aruiz{En este caso, si el número de mascotas que ya tiene registrado el propietario (userPets.length) aún no ha alcanzado el límite establecido en el plan (user.plan.petlimit)}
%\comentario{A. Ruiz}{Este tipo de aclaraciones son fundamentales. las figuras hay que explicarlas} 
This mechanism enables to control the system's execution flow at runtime, facilitating dynamic and adaptive operations without the need for additional code deployments. Such evaluations can also consume complementary data, which is called \emph{toggle context}, to make informed decisions that can depend on the terms of the pricing as shown in figure \ref{fig:togglingExample}. %\comentario{JAParejo}{Aquí lo tenemos a huevo, es el userPets.Lenght}. This information can be allocated in other project subsystems, e.g. the list of pets of an user in Figure \ref{fig:togglingExample}, so it is needed to retrieve it before evaluating the condition. %\agarcia{He eliminado la definición de toggle router porque no aporta nada al ejemplo de la Figura \ref{fig:togglingExample}. Dejo el texto comentado por si lo queremos reutilizar en la sección de los challenges}

%Inside a toggle point the \verb|toggleRouter| acts as an intermediary in the evaluation process, interpreting the toggle configuration, i.e. the declaration of how feature toggles have to be evaluated in order to determine if a feature is available or not, to produce a set of evaluated logical conditions, available through the \verb|featureIsEnabled| method of the example. This mechanism enables selective activation or deactivation of features within the code base at runtime, facilitating dynamic and adaptive operations without the need for additional code deployments. Additionally, a toggle context can be specified inside the router. In short, it provides the complementary data, that usually come from other project subsystems, required by the toggle router to make informed decisions. Inside the SaaS context, this information would include the current user plan, the system's state, or his past interaction with functionalities.

% \begin{figure}
%     \centering
%     \begin{minted}{javascript}
%         function renderAddNewPetButton(){
%             const userPets = fetchUserPets();            
%             if ( userPets.length < user.plan.petLimit ){
%                 return (<NewPetButton/>);
%             }else{
%                 return (<EmptyComponent/>)
%             }
%         }
%     \end{minted}
%     \caption{Caption}
%     \label{fig:enter-label}
% \end{figure}

In order to support the challenges and opportunities of this paper on the basis of an example, we have designed a pricing for PetClinic, a sample veterinary clinic management system designed to help developers to illustrate the functionality and features of a particular software framework or technology, serving as a practical example to understand how to implement various aspects of a technology in a real-world scenario\footnote{The base version of PetClinic for Spring can be found \href{https://github.com/spring-projects/spring-petclinic}{here}}. In our implementation of the system\footnote{Our extended version of PetClinic can be found \href{https://github.com/isa-group/petclinic-react}{here}}, the base functionality has been modified to be driven by the pricing showed in Figure \ref{fig:petclinicPricing}, which comprises 10 distinct features, allocated, restricted or limited across different plans. It also offers subscribers the option of enhancing their experience with additional add-ons that, by rising the cost, provide access to exclusive features not covered in the standard plan offerings. In addition, we have refined the user role hierarchy to incorporate an additional role: \emph{the clinic owner}, who oversees one or more clinics and choose from a variety of plans for each.

\section{Limitations of Traditional Approaches}
\label{sec:limitations}

%\japarejo{para mí es que en realidad la sección 2.1 es en parte el related work, si se amplía un poquito y se habla de los autores que han estudiado las aproximaciones clásicas para los feature toggles dinámicos y los problema que generan, y los que han tratado los planes de precios desde una perspectiva meramente económica creo que ya lo tendríamos prácticamente hecho en la sección 2.1.}

Experience in developing these multi-tenant SaaS systems has shown that establishing a pricing structure that segments users into different roles leads to a significant increase in the variability of the system \cite{GARCIA23}. Besides, as presented by \cite{GHADDAR2012}: ``the real beneﬁt of SaaS begins when it is possible to host multiple customers on the same application instance, without the need of a dedicated application to be deployed and separately maintained for each customer''.

In this scenario, using feature toggles during SaaS development emerges as a popular approach within the community \cite{TOGGLES1,TOGGLES2,TOGGLES3}, since they allow to implement dynamic adaptations of the application state in runtime. By using this approach, the implementation of Petclinic's operational strategy for delivering different User eXperience (UX) configurations could potentially encompass the following elements: i) six feature toggles at the User Interface (UI) level that enable control over the inclusion of distinct features, such as: appointments calendar, vet selection, pet dashboard, online consultations, smart clinic reports and adoption centre; ii) their corresponding toggles within the business logic layer to manage access to these functionalities at the API level, plus iii) two additional dedicated to monitor the constraints set by different plans, such as the maximum number of pets allowed or the number of visits for each one. Finally, iv) a last one will be also needed to manage the support priority to which an user has access to, which is checked at CRM level.

This solution enables Petclinic to offer different UX to pet and clinic owners within a single deployment of the application. Given this framework, one might wonder: how can a new plan be added to the pricing structure?, how much time would it take to change the limitation of allowed pets of the PLATINUM plan? With the current strategies, it would be necessary to expend time in manually modify various \emph{toggle points} throughout the application (technical debt), which significantly increases the chances of making mistakes that might not be covered by the existing suite of tests, and that can lead to catastrophic situations \cite{KNIGHTCAPITALGROUP} 

As can be seen, even Petclinic is a small project, at least 15 \emph{feature toggles} 
%\aruiz{comentar brevemente de donde sale este número} 
are needed to offer three different UX within a single deployment, i.e one for each plan, considering that ``Pets per owner'' and ``Max visits'' do not have any appearances in the front-end and that the ones do have only appear once. Imagine the complexity of managing SaaS with more than 50 features, extra plans and complex add-ons, such as GitHub, Postman, Jira, etc. See \cite{GARCIA23}. 

\section{Challenges of Pricing-driven DevOp of SaaS 
%on the Cloud
}
\label{sec:challenges}
%\japarejo{Figura con los elementos de la arquitectura. y cómo se reflejan los desafíos.}

To design a solution that eases this process, it will be necessary to face the following challenges:

\paragraph{Challenge 1: Pricing-driven feature toggling}

\begin{figure}[htb]
    \centering
    \includegraphics[width=1\textwidth]{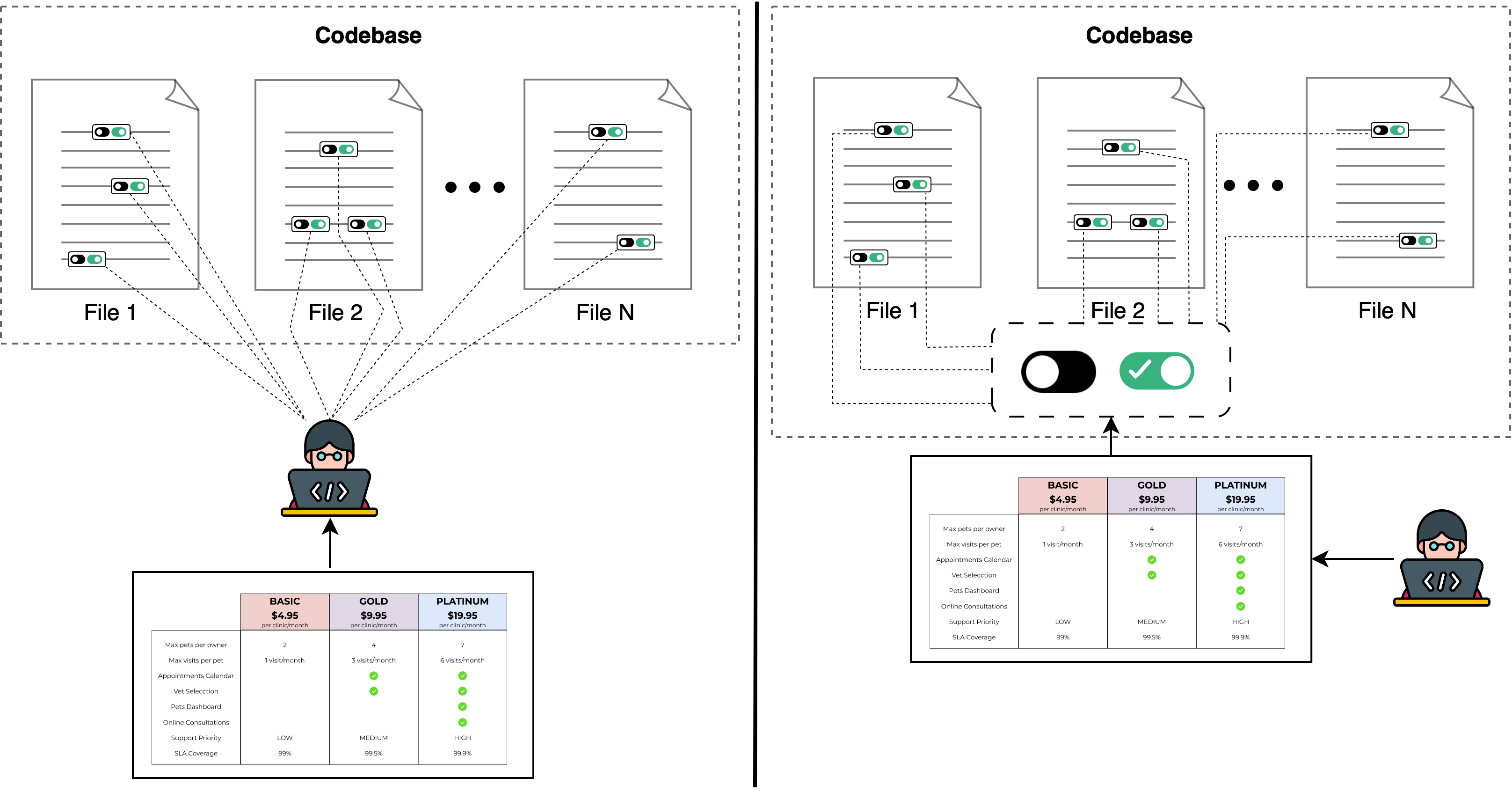}
    \caption{Manual effort required to modifying every affected toggle point versus a centralised pricing-driven toggle router %\agarcia{Hay que modificar el caption}
    %\japarejo{Cambia el pricing para ponerlo a la derecha del desarrollador y más pequeño y ganaremos mucho espacio}} \agarcia{Hecho!}
    }
    \label{fig:challenge1}
\end{figure}

% Los planes de precios se crearon para adaptar un sistema de información a las necesidades de un número mayor de usuarios, de tal forma que cada uno fuese libre de contratar el conjunto de funcionalidades que mejor le viniese.
A primary challenge in SaaS is managing the toggling of features based on the user's subscription level, while avoiding the trap of complex and unmanageable code. As set by \cite{Sun2008Software}, SaaS vendors must create a balance between standardisation and customisation, which involves designing systems that allow for self-adaptive configuration by customers without needing individual code modifications. Such an approach ensures that while the core functionality of the application remains consistent, it can dynamically adapt to diverse plans that can evolve along time without losing code maintainability and readability. Unfortunately, managing these toggles is not that simple, and typically require labor-intesive manual efforts to update every affected toggle point within the application, as illustrated in Figure \ref{fig:challenge1}. This can significantly obfuscate the control flow of the actual business logic of the system, generating technical debt, and complicating testing and quality assurance. For example, as seen in PetClinic, a pricing with 10 features results in, at least, 15 different feature toggles that must be maintained. In addition, mismanagement of these toggles can lead to catastrophic problems. For instance, in 2012, the Knight Capital Group, an American global financial services firm, went to bankrupt due to a repurposed feature toogle that activated dead code that had been unused for 8 years \cite{KNIGHTCAPITALGROUP}.

In order to address this challenge, we propose a centralised pricing-driven toggle router (see Figure \ref{fig:challenge1}) that acts as an intermediary in the evaluation process of every toggle point that consumes the pricing. In this approach, the pricing, in addition to its features, include the declaration of how they must be evaluated in order to determine their availability, producing a dynamic set of evaluated logical conditions that can be used within the toggle points. Besides, the developer can apply changes in the pricing, e.g add a new plan, assuring that the system will automatically adapt to them without requiring manual intervention.
%\japarejo{quizás aquí podríamos apoyarnos en el trabajo relaciona y sus afirmaciones de que la deuda técnica de los feature toggles es de las peores, los proglemas en el equipo de google que mantenía las features experimentales,  y en los efectos catastróficos que se ha visto que ha tenido esto en el pasado.} \agarcia{Hecho!}

\begin{figure}[htb]
    \centering
    \includegraphics[width=1\textwidth]{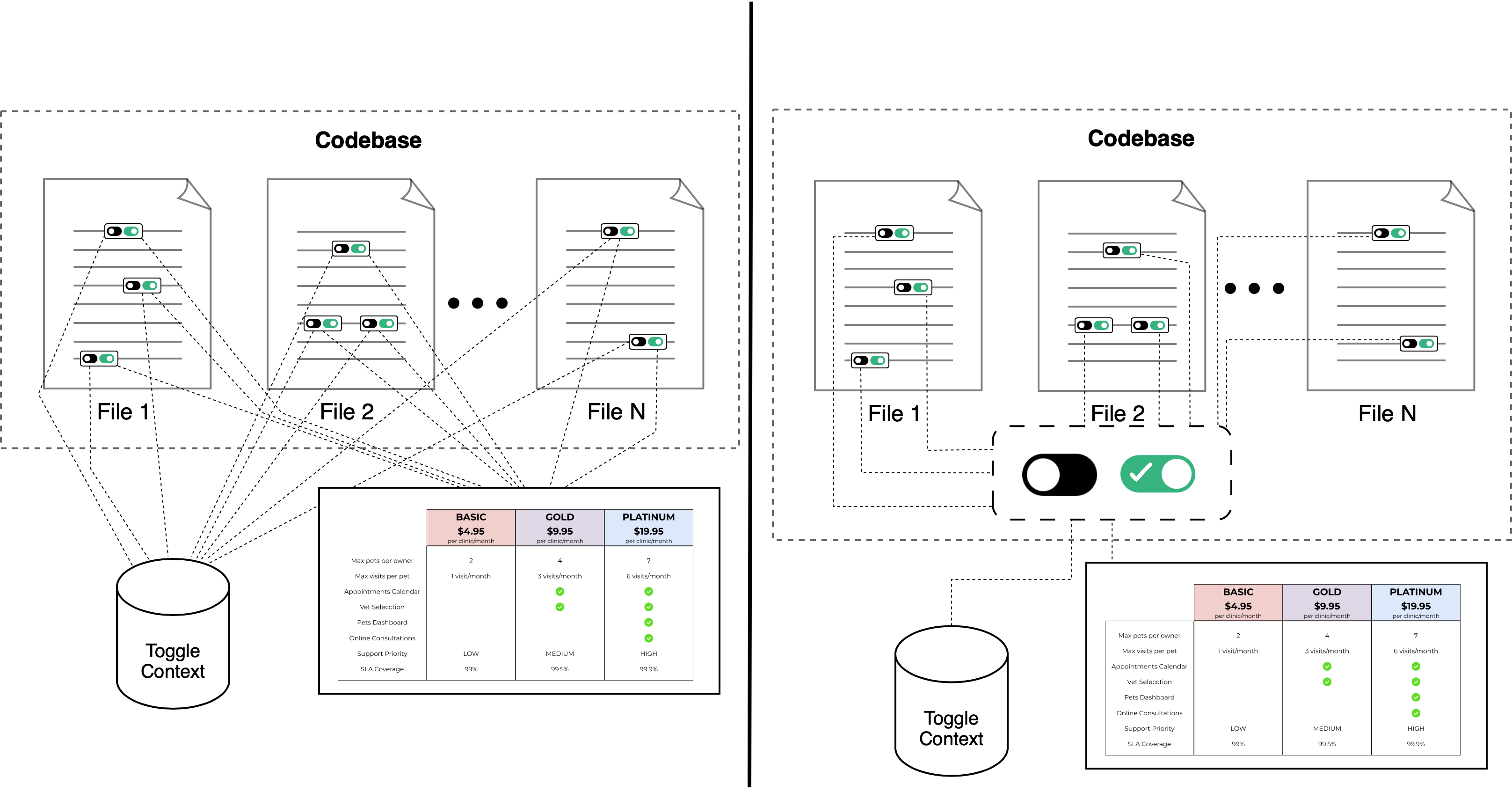}
    \caption{Retrieving data from toggle context and pricing on each toggle point (left) versus using a router for fetching the data needed to all evaluations once (right)
    %\japarejo{de nuevo jugando con el tamaño del pricing podemos ganar espacio vertical recuerda que son 10 páginas}
    }
    \label{fig:challenge2}
\end{figure}

\paragraph{Challenge 2: Stateful and dynamic pricing-drive feature toggling}
%featuSupport dynamic toggling of features and business logic depending on pricing structure and toggle context} 
%\aruiz{el nombre del desafio no debería tener más de 3-4 palabras. ceo que toggling dinámico no se ha presentado en el background, la 'estructura' del pricing no queda claro qa que se refiere. }
In a system with many features to enable or disable depending on the user's plan, the toggling logic usually depends on a toggle context, which defines the complementary data required to make decisions, e.g. the user current usage of functionalities. To evaluate these dynamic and context-dependent toggles at runtime, the system use to hinge on both, a database (DB) that contains a toggle context with information about the status of the system and the user’s past interaction patterns during the subscription period, and the pricing-imposed conditions, such as usage limits or feature availability (Figure \ref{fig:challenge2}).  For example, the ``Pets per owner'' feature of PetClinic requires to extract the current number of pets of an owner from the DB and compare it to the limit imposed by the pricing to the user's subscription in order to determine if the ``Add new pet'' button is displayed or not.
%which depend on the DB state of the application and the user's previous behaviour regarding the application usage during the time the subscription (or other time period) affects. 
Moreover, the challenge becomes even more complex in a distributed environment with client and server logic or micro-services architectures, where the application state is distributed and can be modified concurrently.

Similarly to the first challenge, by using a toggle router within the pricing-driven feature toggling architecture, we can fetch all the data needed by the toggle points of the system from the toggle context and the pricing with a single fetch, significantly reducing their workload.
%using several micro-services.
%This is also particularly difficult in a distributed environment with client and server logic.
%\japarejo{Poner ejemplo con PetClinic, usar las propiedades dinámicas por ejemplo el alta o visitas de mascotas}\agarcia{He propuesto una reescritura, ya me dices ;) (tu texto está comentado aun así). También he añadido el ejemplo}
% El problema no es solamente que existan bastantes características a activar o desactivar, el problema es que estas pueden depender incluso del propio estado de la aplicación, puesto que los planes de precios pueden introducir condiciones dinámicas, que dependan del estado en BD de la aplicación y del comportamiento previo del usuario respecto del uso de la aplicación que se ha realizado durante el tiempo al que afecta la subscripción (u otro periodo de tiempo). Esto es además especialmente difícil en un entorno distribuido con lógica en cliente y en servidor.

\begin{figure}[htb]
    \centering
    \includegraphics[width=0.75\textwidth]{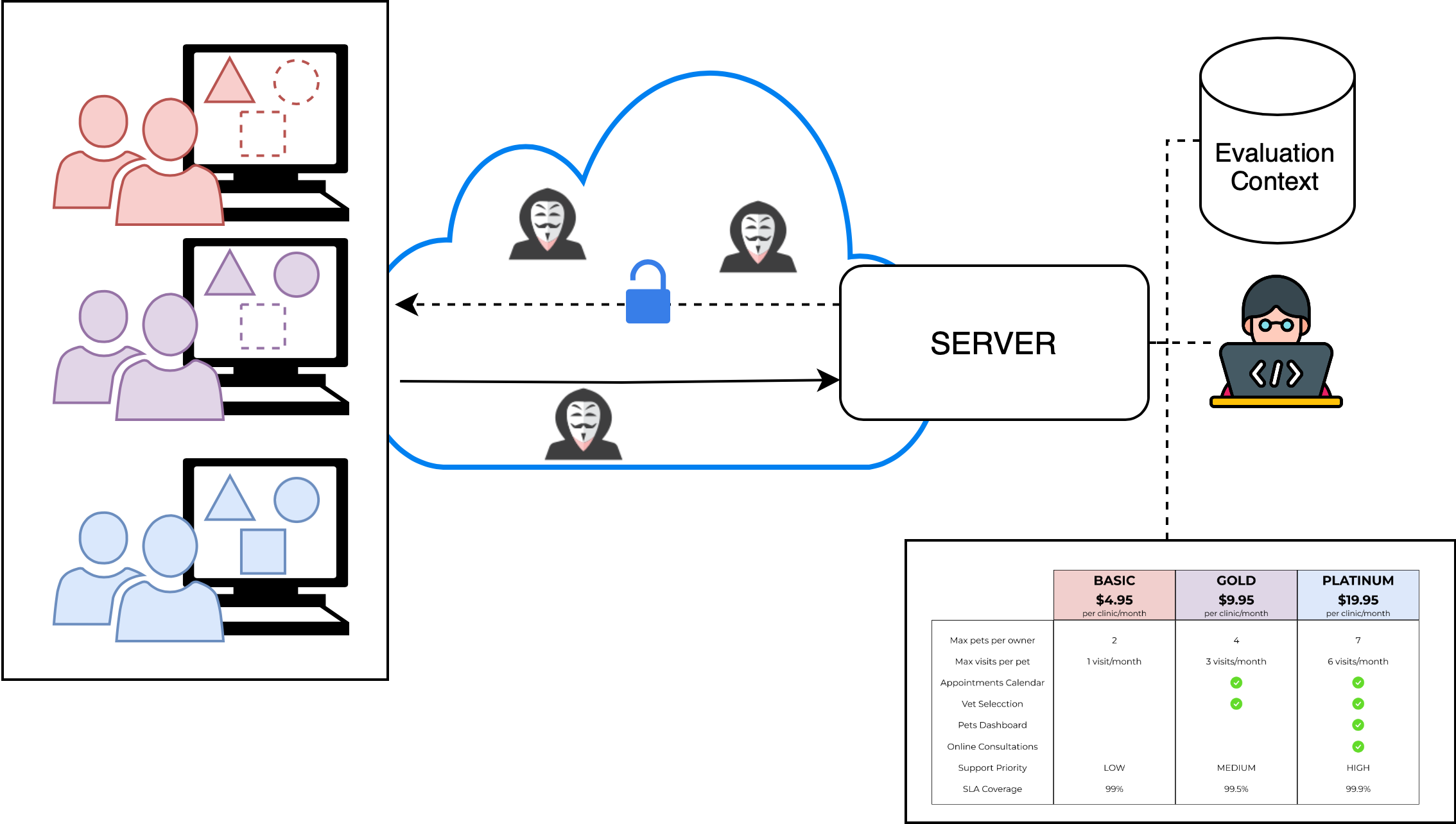}
    \caption{Insecure  %and efficient 
    transmission of feature toggling status through Internet} %\aruiz{resulta confuso que en el caption se hable de eficiente, ¿como transmite eso la figura? de nuevo el marco de los tres tipos de usuarios se puede reducir la intensidad para que o destaca tabnto}
    \label{fig:challenge3}
\end{figure}

\paragraph{Challenge 3: Secure, Unalterable, and efficient transference of subscription status.} %\aruiz{yo lo llamaría 'transfer the subscription status.' y previamente daria un ejemplo del estado, aunmque sea parcial, de la suscripción. simplmente por acortar el 'titulo del desafio' pues despues se desarrolla} 
It is necessary not only to transfer the feature toggling status associated to the user subscription securely, which could be achieved by using secure communications protocols, but also to enable the verifiability and immutability of the subscription status on the fronted, protecting the system from tampering (Figure \ref{fig:challenge3}). Additionally, if the transference of the information requires additional requests from front-end to back-end  with a high frequency,  it will impact strongly on the performance and operational costs of the system. For instance, in PetClinc for the management, since we have a dinamic and stateful feature toggle for allowing the registration of a new pet from the listing of pets, it would be necessary to update the state of the toggles on fronted to know whether or not to show the add toggle button (or create an interface with a poor user experience that lets you click on things only to then tell you that you can't do them). This usually involves a dedicated API request to update the toggling evaluation context, generating overhead and inscreased response times. %\japarejo{Creo que desde la perspectiva de la eficiencia si se podría poner un ejemplo en el contexto de pet-clinic, por ejemplo que en el caso de la visualización del listado de mascotas, al tener una  feature con toggling dinámico, habría que actualizar el estado de los toggles para saber si se debe mostrar o no el botón de añadir toggle (o crear una interfaz con una experiencia de usuario pobre que te deja pulsar en cosas que luego te dice que no puedes hacer), y esto sería para cada vez que se mostrar el listado!.}
%For example, a PetClinic's client must not be able to modify the evaluation of a feature to make it accessible illegally.

\paragraph{Challenge 4: Monitoring and enforcement of pricing plans usage constraints and guarantees (including QoS-based or SLA-related ones)} 
% Aún hoy en día, es un desafío  dimensionar una arquitectura cloud para que se genere el cumplimiento de las garantías de un SLA único para todos los clientes, en una arquitectura donde dependiendo existen distintas condiciones y garantías dependiendo de la subscripción del usuario, y con un volumen de usuarios diferentes asociados a cada plan, el problema se vuelve aún más complejo.
Even nowadays, scaling a cloud architecture to achieve compliance with the guarantees of a single SLA for a wide set of customers is challenging. So, in an architecture where there are different conditions and guarantees depending on the user's subscription, and in scenarios with a different volumes of users associated with each plan, the problem becomes even more complex.
%\japarejo{Creo que no hace falta ejemplo porque es bastante evidente}\aruiz{eso lo dices porque tu sabes mucho ;-). creo que hace falta. En cualquier caso, si no entrá en el ámbito de Pricign4SaaS, quizás se pueda obviar. así hay más espacio para poder desarrollar cosas que puedan ser necesarias en secciones anteriores, sobretodo, teniendo en cuenta la segunda versión que enviaremos en algún momento del futuro. este finde o en el camera-ready}

%\japarejo{Quizás el la oportunidad más importante es la 2 (ahora es la 1) si se alcanza la automatización completa (o lo más completa posible) del princing-driven development and operacion of SaaS, consiguiendo que un cambio en el plan de precios se difunda de la manera más automatizada posible desde su seralizaicón a la lógica que la implementa y la operación del sistema con los depliegues (en la nube, on premise o híbridos) que precise para hacer honor a las condiciones que establece el plan de precios. } \aruiz{pues si}
\section{Opportunities of Pricing-driven DevOp of SaaS}
\label{sec:opportunities}

It becomes clear that this approach to pricing management transcends monetary considerations, influencing the core of SaaS development and operation, and generating a deep impact on market success and customer satisfaction. By addressing the challenges inherent in pricing-driven development of SaaS, new opportunities arise.

%\japarejo{Esto includo oportunidades para aprovechar la elasticidad de las infraestructuras en la nube en base a la información conocida de subscripciones a cada plan y los patrones de uso observados por cada tipo de usuario en los distintos planes. Ejemplo en el caso de pet-clinic, decidir una migración a una infraextructura con un opex menor si se aprecia que vamos sobrados de infrastructura, pero para eso hay que resolver el challenge 4.}

\paragraph{Opportunity 1: Automated pricing-driven feature toggling} 
If we succeed in streamlining feature toggling in SaaS, providers will be able to design and alter their pricing plans more frequently, enabling them not only to keep pace with current market trends but also to potentially set new ones. For example, in PetClinic, we would be able to release new features, such as aesthetic treatments or pet hotel services in the clinic, that are regulated by the pricing right after their implementation and testing, without needing to expend extra development time. The key to successfully implementing these strategies lies in overcoming challenges 1 and 2, which focus around enhancing the management of feature toggles. This opportunity is key to faster market adaption and user customisation.

%\japarejo{Si se consigue agilizar el proceso de adaptación del feature toggling a los cambios de los planes de precios, se podrán diseñar y cambiar los planes de precios con mayor frecuencia ayudando a seguir o incluso a tendencias en el mercado de productos del domino en el que operen los SaaS. Poner un ejemplo en petclinic, por ejemplo introduciendo una nueva condición para tratmientos estéticos o de cuidado/estancia temporal en la clínica por plan. Hay que resolver los challenges 1 y 2 para esto}\agarcia{Hecho!}

\paragraph{Opportunity 2: An actual common language between business decision-makers and developers}

If we could harmonise the vocabulary of the pricing plan with the variables controlling feature toggling, it would enable modifications in the pricing plan that are dynamically evaluated by the toggles without needing to alter their code. This approach not only streamlines the process but also ensures that business strategies and technical implementations are more closely aligned. It creates a more efficient, resilient and responsive system where business decisions can be quickly translated into technical actions, reducing the lag between decision-making and implementation. In our running example, a non-technical user could take the role of PetClinic's administrator and apply changes to the current pricing that are automatically integrated within the system.

%\japarejo{Si pudieramos aunar el vocabulario del plan de precios con el de las variables que controlan el toggling, podríamos hacer que las modificaciones en el plan de precios, debidamente serializado en un formato legible por la máquina, fueran evaluados de manera dinámica por los toggles sin tener que alterar su código (es lo que hemos hecho básicamente).}

\paragraph{Opportunity 3: Improved alignment of deployment infrastructure and its usage based on subscription data} 
This includes leveraging the elasticity of cloud infrastructures, which can be optimised based on known subscription information for each plan and the observed usage patterns of different user types across these plans. For instance, PetClinic could lead to a decision to migrate to an infrastructure with lower operational expenses (OPEX) if it is observed that the current infrastructure provides significantly more processing or storage capacity than actually needed. However, to achieve this, challenge 4 must be addressed.

\section{Conclusions}
\label{sec:conclusion}
%\japarejo{Si vemos que no aportan demasiado las conclusiones y son un mero resumen se pueden quitar directamente para entrar en espacio.}
In this paper, we have explored the problems of managing a pricing-driven SaaS solution on the cloud, using PetClinic as a running example. Our analysis reveals the complexity of maintaining and developing a system that has to manage multiple feature toggles to offer different UX within a single deployment, showing that the future of SaaS lies in the harmonious integration of business models with technological capabilities.

Ultimately, Pricing-driven Development and Operations of SaaS on the Cloud is more than a methodology —it's a strategic choice that reflects the evolving nature of software services in the cloud era. Embracing its challenges is not optional but necessary for SaaS providers aiming to remain competitive and responsive to the ever-changing demands of the market. This paper, therefore, serves as both a guide and a call to action for those navigating the complex, yet rewarding, landscape of SaaS on the cloud.

\subsubsection{Acknowledgements} 
This work has been partially supported by grants \\
PID2021-126227NB-C21, and % PERSEO
PID2021-126227NB-C22      % ATENEA
funded by MCIN/AEI/ \\
10.13039/501100011033/FEDER and European Union ``ERDF a way of making Europe'';
and %
TED2021-131023B-C21 and % IRIS
TED2021-131023B-C22     % ORCHID
funded by MCIN/AEI/10.13039/501100011033 and European Union ``NextGenerationEU''/ \\
PRTR;

%
% ---- Bibliography ----
%
% BibTeX users should specify bibliography style 'splncs04'.
% References will then be sorted and formatted in the correct style.
%
 \bibliographystyle{splncs04}
 \bibliography{references}

\end{document}